\documentclass[prl,twocolumn,showpacs,preprintnumbers,amsmath,amssymb]{revtex4}% Physical Review B

\usepackage{graphicx}% Include figure files
\usepackage{dcolumn}% Align table columns on decimal point
\newcommand{\beq}{\begin{equation}}
\newcommand{\eeq}{\end{equation}}
\newcommand{\beqa}{\begin{eqnarray}}
\newcommand{\eeqa}{\end{eqnarray}}

\newcommand{\qvec}{{\bf q}}

\begin{document}
\title{Electron-phonon interaction in Strongly Correlated Systems}

\author{M. Capone$^{1,2}$, C. Castellani$^{1}$ and M. Grilli$^{1}$}

\affiliation{$^1$CNR-INFM Statistical Mechanics and Complexity Center, 
and Dipartimento di Fisica\\
Universit\`a di Roma ``La Sapienza" piazzale Aldo Moro 5, I-00185 Roma, Italy \\
$^2$Istituto dei Sistemi Complessi del CNR, Via dei Taurini 19, I-00185 Roma, Italy}

\begin{abstract}
The Hubbard-Holstein model is a simple model including both electron-phonon interaction and electron-electron correlations. We review a body of theoretical work investigating the effects of strong correlations on the electron-phonon interaction. We focus on the regime, relevant to high-$T_c$ superconductors, in which the electron correlations are dominant. We find that electron-phonon interaction can still have important signatures, even if many anomalies appear, and the overall effect is far from conventional. In particular in the paramagnetic phase the effects of  phonons are much reduced in the low-energy properties, while the high-energy physics can be strongly affected by phonons. Moreover, the electron-phonon interaction can still give rise to important effects, like phase separation and charge-ordering, and it assumes a predominance of forward scattering even if the bare interaction is assumed to be local (momentum independent). Antiferromagnetic correlations reduce the screening effects due to electron-electron interactions and revive the electron-phonon effects.
\end{abstract}
\date{\today}
\maketitle
\tableofcontents

\section{Introduction}\label{sec:intro}
A wealth of materials, including the most challenging systems (cuprates, manganites, fullerenes, \ldots), present clear signatures of both electron-electron (e-e) and electron-phonon (e-ph) interactions, leading to a competition -or interplay- which can give rise to different physics according to the value of relevant control parameters and of the chemical and electronic properties of the materials. In this topical review we will focus on high-temperature superconductors, with the copper-oxide compounds (cuprates) in a prominent role, and an attention to the alkali-doped fullerides.

In the case of the cuprates, which are arguably the most accurately studied materials in the last twenty-five years, the signatures of electron-phonon interactions are nowadays clear\cite{Gunn_cuprates}, eventhough the overall scenario is far from ordinary: Electron-phonon fingerprints are evident in some properties, while they are weak or absent in other observables.
Specifically, clear polaronic features are observed in optical conductivity\cite{polaronexp} as well as in angle-resolved photoemission experiments (ARPES)\cite{KMShen} in very lightly doped compounds. A substantial e-ph coupling can also be inferred by the Fano line shapes 
of phonons in Raman spectra and by the rather large frequency shift
and linewidth broadening of some phonons at $T_c$. Phonons are also good candidates to
account for the famous ``kink'' in the electronic dispersions observed
in ARPES experiments\cite{Lanzara,reznik2}.  On the other hand phonons, which typically affect resistivity
in standard metals, do not appear in transport experiments on cuprates. For instance the resistivity
around optimal doping is ubiquitously linear in temperature (even in systems with relatively low critical temperature) and no high-temperature saturation seems to
be present up to the highest achieved temperatures. This dichotomic behavior of cuprates, 
which display clear phononic features in some cases and none in others is one of the puzzling and challenging
issues raised in these materials. Yet, it is almost universally recognized that the main player in the cuprate game is the e-e correlation, which drives the parent compound insulating and it is probably responsible of the main features of the overall phase diagram, including d-wave superconductivity and the existence of a pseudogap above the superconducting critical temperature. Therefore, it is not surprising that the signatures of e-ph interaction in the cuprates can hardly be understood in terms of the standard theory of e-ph interactions in weakly correlated metals, and a new theoretical framework including e-e correlations is needed. We will argue here that this change of perspective can indeed reconcile the different relevance of phonons in the various observables in correlated systems. 

On the other hand  the superconducting members of the fulleride family, of composition A$_3$C$_{60}$ with A an alkali-metal atom, are often considered as standard phononic superconductors, in which the coupling between electrons and the local vibrations associated to the distortion of the carbon bukyballs is the driving force of superconducting pairing. The conventional nature of these compounds is challenged by recent investigations in expanded fullerides revealing several physical properties associated to e-e correlations. Indeed the Cs$_3$C$_{60}$ solid with A15 structure is an antiferromagnetic Mott insulator at ambient pressure which becomes superconducting only under applied pressure, with $T_c$ reaching 38K\cite{C60_corr}. The phase diagram as a function of pressure closely resembles that of cuprates as a function of doping, suggesting a central role of correlations. Consequently, e-e interactions are expected to be important in other members of the fulleride family. Indeed it has been shown that, thanks to the orbital degeneracy and the Jahn-Teller nature of the relevant phonons, there is no contradiction between a phonon-mediated superconductivity and the relevance of electronic correlations, and the two interactions turn out to cooperate in providing relatively high critical temperature\cite{capone_science}. 

In an extremely broad sense these materials (cuprates and fullerides), as well as many others that we did not talk about, raise the same conceptual problem, namely the investigation of systems in which both e-e interaction and e-ph coupling are non negligible and the physics can be explained only taking both into account. On the other hand the same phenomenology suggests that this competition may result in completely different physics according to specific aspects of the materials. 
In general we can expect different behaviors because of: {\it (i)} Different parameters within the same model (e.g., which is the largest scale between electron-phonon interaction and electron-electron repulsion); {\it(ii)} Different form for the interaction term, or more generally, different models.

If we focus on point {\it (i)} we immediately realize that even the simplest models must contain several relevant physical parameters. As we will discuss in the following, we have to deal at least with the electronic bandwidth, the Coulomb repulsion, the strength of electron-phonon interaction, the phononic frequency and the chemical potential that controls the band filling. This determines a multi-dimensional phase diagram, which can hardly be understood in its entirety within a single analysis and it is expected to present several different regimes.
Therefore, even if we choose one given simple model, it may be useful to focus on a given physical regime, which essentially implies to select a hierarchy between the different energy scales, or to fix (or neglect) some of them.

Our choice, is to focus on the ``strongly correlated" metallic phases, i.e., on system in which the Coulomb repulsion is the largest energy scale. The polar star of this work is the understanding of the fate of electron-phonon interaction in systems that are dominated by electron-electron interactions such as the cuprates. Nonetheless, our discussion will also follow some detours, which will help us to build a more comprehensive picture of the competition beteen the two interactions. One of these detours will touch point {\it (ii)} addressing the role of the phonon symmetry in its interplay with correlations. This point is crucial for the understanding of the sinergy between e-ph interaction and e-e correlation in the fullerenes.

The manuscript is organized as follows: In Sec. II we introduce the Hubbard-Holstein model. Sec. III is devoted to a Fermi-liquid analysis of the effects of correlations on electron-phonon interactions and to a mean-field solution of the Hubbard-Holstein model within the slave-boson formalism. Sec. IV presents a non-perturbative Dynamical Mean-Field Theory study of the Hubbard-Holstein model.  Sec. V is dedicated to the charge instabilities of the model; Sec.VI briefly compares the Hubbard-Holstein model with a three-band model with Jahn-Teller interactions introduced for the fullerenes. Sec. VII presents our conclusions.

\section{The model}
\label{sec:model}
The simplest model of a strongly correlated electron system coupled to the lattice is given by the single-band Hubbard-Holstein (HH) model
\begin{eqnarray}
H & = & -t \sum_{\langle i,j \rangle , \sigma} 
\left( c^\dagger_{i\sigma} c_{j\sigma} + H.c.\right) \nonumber \\
%-t' \sum_{\langle \langle i,j \rangle \rangle , \sigma} 
%\left( c^\dagger_{i\sigma} c_{j\sigma} + H.c.\right)\nonumber \\ 
& - &\mu_0\sum_{i\sigma} n_{i\sigma}
  +  U\sum_i n_{i\uparrow}n_{i\downarrow} \nonumber \\
& + & \omega_0 \sum_i a^\dagger_i a_i + g \sum_{i,\sigma}
\left( a^\dagger_i+a_i\right) \left( n_{i\sigma} -\langle n_{i\sigma} \rangle 
\right) ,\label{HHHam}
\end{eqnarray}
where $\langle i,j \rangle$ 
%and $\langle \langle i,j \rangle\rangle$ 
indicate nearest-neighbor 
%and next-nearest-neighbor 
sites.
$n_{i\sigma}=c^\dagger_{i\sigma} c_{i\sigma} $ is the local
electron density, which is coupled via $g$ to the field $a_i$
of a dispersionless phonon. This model is clearly a very idealized description of a solid: both the Coulomb repulsion and the electron-phonon interaction are assumed to be purely local, only one electron band is considered, only the electron density couples to phonons. 
The relevant physical parameters are the strength of the coulomb
repulsion $U$, the dimensionless electron-phonon coupling $\lambda =
2g^2N_0/\omega_0$ (here $N_0$ is the electronic density of states at the
Fermi level), the adiabatic ratio $\omega_0/W$ ($W$ being the electronic
bandwidth), and the value of the chemical potential and the details of the bandstructure (inclusion of next-neighbor hopping). The dimensionality of the system also plays a major role.
This multi-dimensional parameter space leads to an extremely rich physics, and the number of paper devoted to this simple model is countless. Various approaches have been considered to solve the HH model in the presence of strong correlations. Besides numerical techniques like quantum Monte Carlo,~\cite{hir83, Hir85, Ber95, Hua03, korni} exact diagonalization,~\cite{stephan,fehske,ranninger}  Dynamical Mean Field Theory (DMFT),~ \cite{FJ95, Capo04, Kol04-1, Kol04-2, Jeo04,San05, San06,werner,paci,macridin}, Density-Matrix Renormalization Group~\cite{dmrg}, (semi)analytical approaches like slave bosons (SB) \cite{GC,BTGD,Kel95,Koc04,Cappelluti_sb}, large-N expansions \cite{KZ1,KZ2}, variational approaches\cite{napolivar} including a modified Gutzwiller scheme\cite{barone1,barone2,DLGS}, have been useful to elucidate the renormalization of the e-ph coupling in the presence of (strong) correlations.

According to our choice, in this work we will mainly consider on the large-$U$ regime, and on regions of parameters in which the system is a strongly correlated metal. Even if our focus will be the strongly correlated HH model, we will discuss its results in comparison with some related models, like the Hubbard-tJ model, the three-band Hubbard model for the cuprates and a three-orbital Hubbard model with Jahn-Teller interactions for the doped fullerides.
Our investigation will be mainly dedicated to the effects of e-ph interaction on the self-energy and the quasiparticle renormalization factor $z$ starting from a strongly correlated regime, in which the Coulomb interactions is strong enough to determine the physics of the problem, but the system is not a Mott insulator.
Since the study of superconductivity in DMFT (our main tool of investigation) is limited to s-wave symmetry we will not discuss superconductivity in the HH model which is expected to be d-wave if correlation dominates. On the other hand we will study s-wave superconductivity in a three-orbital model for fullerides which emphasizes the role of the symmetry of e-ph interaction in presence of correlations.

\section{Effect of electron-phonon interaction in a correlated metal}
\label{sec:effect}
\subsection{Fermi-liquid analysis}
\label{sec:fl}
In this section we begin our analysis of the properties of e-ph interaction in a correlated metal within a Landau Fermi-liquid (FL) picture\cite{Nozieres,AGD}.
Within this approach, the correlated metal is described as a collection of quasiparticles with an effective mass $m^*$ instead of the physical electron mass $m$. In the presence of strong e-e correlations the motion of the carriers is naturally obstacled by the interactions, which is reflected in a large ratio between the effective mass and the bare mass $m^*/m \gg 1$ and in a loss of low-energy spectral weight described by a small quasiparticle renormalization factor $z_e$. The former reflects in an enhanced quasiparticle density of states $N^* = m^*/m N_0$ ($N_0$ being the bare density of states) and the latter renormalizes the quasiparticle interactions.

In order to characterize the fate of the e-ph interaction in a similar correlated metal we need to consider also the vertex corrections introduced by e-e interactions, for which no Migdal theorem can be invoked.

We can gain a first insight on the way in which the e-ph interaction behaves in the presence of strong correlations by considering
the effective dimensionless e-e interaction mediated by the exhange of a single-phonon
\beq
N^*\Gamma_{eff}^{ph}(\qvec,\omega)= N^* g^2 {z_e}^2{\Lambda_e}^{2}(q,\omega) \frac{2\omega(q)}{\omega^2-\omega^2(q)},
\label{geffoneph}
\eeq
by assuming that all the correlation effects are local, or equivalently that the self-energy is independent on momentum, the effective mass is related to $z_e$ by  $z_e = m/m^*$.  In Eq. (\ref{geffoneph}) ${\Lambda}_e$ includes the vertex corrections which renormalizes the e-ph vertex, and the last factor is the free phonon propagator. In order to focus on the effect of correlations on the phonon-mediated interaction, in both $z_e$ and ${\Lambda}_e$ we include only processes due to e-e interactions, as reminded by the index ``e'' that we attached to them.

Within a Landau Fermi-liquid picture we can use  the Ward identities that connect the vertex corrections $\Lambda$ with the wavefunction renormalization $z$. In the small frequency ($\omega$) and transferred momentum ($q$) regimes these identities have two distinct forms depending on the order of the 
$\omega \to 0$ and $q\to 0 $ limits. In the case of the charge-density vertex, which is relevant for our Holstein coupling, we have
\beqa
z_e{\Lambda}_e(\omega\to 0, q=0)&=&1 \nonumber \\
z_e{\Lambda}_e(\omega= 0, q\to0)&=&\frac{1}{1+F_0^s} 
 \label{wi1}
\eeqa
where $F_0^s$ is the symmetric Landau parameter. (\ref{wi1}) are exact Ward identities, which are satisfied irrespective of the details of the e-e interactions
and show the drastic difference between the dynamic [$(\omega\to 0, q=0)$] and static [$(\omega= 0, q\to 0)$] limits.%Whenever the exchange of a phonon takes place, the vertex corrections must be included leading to a different behavior of the effective interaction in the two limits. 

%where $\omega(\qvec)$ is the phonon dispersion. Here the presence of $z_e$ indicates that we are considering the interaction between quasiparticles and $\Lambda^e$ expresses the difference of the phonon coupling to the quasiparticles with respect to particles \cite{Nozieres}.
%The effect of the strong $\omega$ and $\qvec$ dependence of $\Lambda^e$ in $\Gamma_{eff}$ is made apparent in the small $\qvec, \omega$ limits, where the relations (\ref{wi1}) can be used. Then one obtains
Plugging these results into Eq. (\ref{geffoneph}) we obtain, in the two limits considered above
\beqa
N^*\Gamma_{eff}^{ph}(\omega \to 0, q = 0) &=& -\frac{2g^2N^*}{\omega_0} \\
N^*\Gamma_{eff}^{ph}(\omega = 0, q \to 0) &=& -\frac{2g^2N^*}{\omega_0}\frac{1}{\left[1+(F_0^{s})_e\right]^2}=\nonumber \\
&=& -\frac{2g^2}{\omega_0}\frac{{\kappa^e}^2}{N^*},
\label{effgam}
\eeqa
where $\kappa^e=N^*/[1+(F_0^{s})_e]$ is the charge compressibility in the absence of e-ph coupling.
The difference between the dynamic and static case can be dramatic in the case of a Fermi liquid
with a large mass enhancement $m^*/m \gg 1$, and small compressibility renormalization
$(\kappa^e \sim N_0)$. This requires $(F_0^{s})_e$ to be much larger than one and proportional to the quasiparticle density
of states $N^*=(m^*/m)N_0\gg N_0$. Eq. (\ref{effgam})
then leads to 
\beqa
N^*\Gamma_{eff}^{ph}(\omega \to 0, \qvec= 0) &=& -\lambda \left(\frac{m^*}{m}\right)  \label{prima} \\
N^*\Gamma_{eff}^{ph}(\omega =0, \qvec \to 0) &=&  -\lambda \left(\frac{m}{m^*}\right)\label{seconda}
\eeqa
so that the effective one-phonon mediated e-e interaction is large $(\sim m^*/m)$ in the 
dynamic limit and small $(\sim m/m^*)$ in the static one. We emphasize that the key condition for the equalities
(6,7) to hold is that $\chi^e \ll m^*/m$.
Therefore they are verified also for a modest mass enhancement as long as the compressibility remains much smaller than it.
 
The strong $\omega-q$ dependence in Eqs.(6,7) has been demonstrated on really general grounds 
only in the small-q and small-$\omega$ limits, whereas the case of finite $q$'s 
and $\omega$'s needs (approximate) analyses of specific models. 
The cases of the single-band
and of the three-band Hubbard models with infinite local repulsion
have been extensively considered in the literature
as prototypical models of strong correlations. In this framework the issue of e-ph coupling
has been considered by means of the Holstein \cite{BTGD,DLGS,KZ1,GC,San05}
or (less frequently) of the so-called Su-Schrieffer-Heeger
coupling (where phonons couple to the electron hopping term) \cite{KT}.

Results for these models show that the product $z\Lambda$
remains of order one in the dynamical regime as long as the momentum and frequency lie outside the 
particle-hole continuum, while it is strongly suppressed (as in the static limit) inside the particle-hole continuum, where important
screening processes take place. Moreover, strong correlations 
provide further suppression of the e-ph coupling when, within the static limit $\omega=0$, the 
transferred momentum is increased \cite{KZ1}. These additional screening  channels 
depend on the details of the electronic band structure determining particle-hole screening processes.

The general Fermi liquid discussion and the specific analysis of models with strong correlations
generically demonstrates the relevant role of dynamics in the screening effects that e-e correlations
induce on the e-ph coupling. This strong dependence of the e-ph vertex on momentum and frequency (and on their ratio) 
makes the effects of the e-ph coupling rather subtle, since different physical quantities,
involving different dynamical regimes, may display more or less suppressed e-ph effects.
In particular the e-ph coupling (and the e-e interaction mediated by phonons) will be depressed
by strong e-e interactions whenever small energy and large momentum transfer are involved
(like, e.g., in transport). This suppression may be substantial, for instance, in the low-doping region
of the superconducting cuprates, where e-e correlations are strong due to the relative proximity to
a correlation-induced insulating phase. On the other hand different physical processes involving
dynamical processes could experience a more pronounced e-ph coupling.
Specific calculations carried out in a single-band Hubbard-Holstein model within a large-N
treatment of the e-e correlations \cite{KZ1,KZ2} demonstrated that the Eliashberg spectral function
$\alpha^2F(\omega)$ determining superconductivity is much less reduced than the analogous 
quantity $\alpha^2F_{tr}(\omega)$ determining transport properties.
As we will discuss in the following, this different renormalization will found a counterpart in nonperturbative
dynamical mean-field theory calculations.

Even if our focus is on the Hubbard-Holstein model, it can be useful to recall that in the case
of phonons coupled to the electron current, Ward identities similar to those of Eq. (\ref{wi1})
can be derived \cite{Nozieres}
\beqa
z\Lambda_\alpha(\omega\to 0, q=0)&=&J_{q\alpha}  \nonumber \\
z\Lambda_\alpha(\omega= 0, q\to0)&=&v_{q\alpha} 
 \label{wi2}
\eeqa
where $\Lambda_\alpha$ is the $\alpha-th$ component of the electronic vector vertex part. 
$v_{q\alpha}$ and $J_{q\alpha}$ are instead the $\alpha$-th components of the quasiparticle
velocity and of the current respectively. Similarly to the case of a coupling
with the density, the correlations suppress much more strongly the e-ph coupling
in the static limit, while they affect little the
coupling in the dynamical case (remember that the current is weakly touched by interactions
and it even remains constant in translationally invariant systems).

\subsection{The Hubbard-Holstein model: mean-field slave-boson approaches}
\label{sec:sb}

A natural tool to adreess the screening effects beyond the small-q and small-$\omega$ regime 
are mean-field approaches based on a slave-boson representation of the Hilbert space.
These methods, although approximate, capture the main physical ingredients
of the problem with a description of a Fermi liquid of quasiparticles coupled
by a residual interaction. This is why we briefly summarize here
some results obtained \cite{BTGD} in the simplest formulation of 
the slave-boson large-N approach to the infinite-U Hubbard model.
Here one can reach a semiquantitative understanding of the effects of
correlations on the e-ph coupling, which are in substantial agreement
with the results of more sophisticated approaches.

In this discussion we will consider the infinite-repulsion limit, which simplifies
the formalism. In this limit we have a sharp contraint of no double occupancy on 
each lattice site $\sum_\sigma n_{i\sigma} \le 1$.
The standard slave-boson technique implements the constraint
\cite{barnes,coleman,RN,KL,slabos} by performing the usual substitution
$c^{\dagger}_{i \sigma}\rightarrow f^{\dagger}_{i
\sigma}b_i, \,\,\, {c}_{i \sigma}\rightarrow b^{\dagger}_i f_{i
\sigma}$, where the fermionic $f_{i\sigma}$ operators represent
quasiparticles, while the bosonic field $b_i$ labels the
state of a site with no fermions in it. 
This formulation of slave bosons is here used in
connection with a large-$N$ 
expansion \cite{coleman} in order to introduce a small
parameter allowing for a systematic perturbative expansion
without any assumption on the smallness of any physical quantity.
Within the large-$N$ scheme, the spin index
runs from 1 to $N$ and the constraint assumes the form
$\sum_\sigma c^{\dagger}_{i\sigma} c_{i\sigma} +b^{\dagger}_ib_i = 
\frac{N}{2}$. A suitable rescaling of the hopping amplitudes
$t_{ij} \rightarrow t_{ij}/N$ must, in this
model, be joined by the similar rescaling of the $e$-$ph$ 
coupling $g \rightarrow {g/ {\sqrt{N}}}$ in order to
compensate for the presence of $N$ fermionic degrees of freedom.
It is beyond the scope of this review to report the technical details
of this technique, which have been extensively presented in previous 
works \cite{BTGD}. Here we simply remind that the leading order in the
large-N expansion provides a mean-field description of the infinite-U 
Hubbard model with uniform constant values of the bosonic field $b_i \equiv b_0$ and
of the Lagrange multiplier field $\lambda_i= \lambda_0$ implementing 
(on the average) the no-double-occupancy constraint. This
gives rise to an insulating phase at half-filling ($n=1$,
doping $\delta=0$) and a Fermi liquid metallic phase at finite doping
with small quasiparticle residue $z=b_0^2=\delta$ and large mass
$m^*=1/z$. The treatment of the fluctuations at the next leading order beyond mean-field introduces residual interactions
between the quasiparticles and allows to determine the scattering amplitudes 
in the particle-particle (Cooper) channel $\Gamma^C(k,k';q,\omega)$ and
in the particle-hole channel $\Gamma(k,k';q,\omega)$. Taking the small-(q,$\omega$)
limits in this latter quantity also allows to determine the Landau parameter 
\begin{equation}
F_0^s=N^*\Gamma_\omega=4tN^*\varepsilon_{k_F} -\lambda \frac{N^*}{N_0}.
\label{f0s}
\end{equation}
Here $\varepsilon_{k_F}$ is the electron dispersion calculated at  the Fermi energy.
As we already noted, $F_0^s$ enters in the FL expression of the compressibility. When $F_0^s < -1$ the
thermodynamic stability condition $\kappa > 0$ is violated and the system undergoes phase separation.
We shall discuss this issue in Sec. VI. 
We anticipate that, in the phase separated region, long-range
Coulombic forces play a crucial role. It is indeed natural that a long-range 
interaction frustrates the formation of charge-rich regions. The outcome of this 
competition is, as we shall see, a shift of the charge instability
to finite momenta promoting the formation of a charge-density-wave
phase. 
In this section we will consider parameters far from the phase separation instability. Nonetheless,
in light of the central role played by long-range interactions in the phase separated case, we will also 
comment about their effect in these stable regions of parameters.

\subsubsection{Static properties}
\label{subsec:static}
\begin{figure}[t]
\includegraphics[width=5cm,angle=0,clip=true]{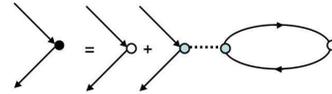}
\caption{Leading-order in $1/N$ diagrammatic structure of the
effective $e$-$ph$ vertex dressed by electronic processes only:
the dashed line is the slave-boson propagator only involving
$b$ and $\lambda$ bosons (pure e-e interaction), the solid dot is the dressed $e$-$ph$
vertex, the open dots are the bare $e$-$ph$ vertices and
the grey dots are the quasiparticle-slave-boson vertices. 
}
\label{diagramG}
\end{figure}

\begin{figure}[t]
\includegraphics[width=7cm,clip=true]{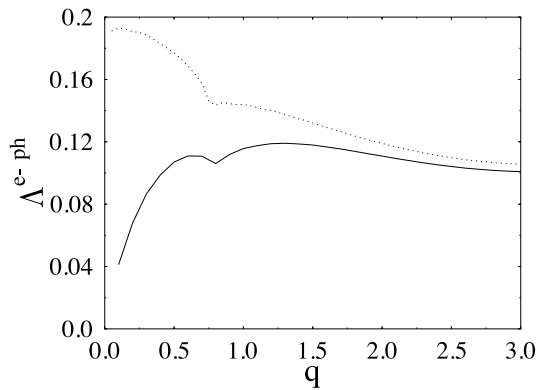}
\caption{Static effective $e$-$ph$ vertex in units of the bare e-ph coupling $g$
as a function of  the transferred momentum (in units of the inverse lattice spacing $1/a$) 
in the (1,0) direction. The vertex is calculated at leading order in $1/N$ 
 for a HH model with $t = 0.5$eV, $t'=-(1/6)t$, $\omega_0=0.04$eV and doping 
$x=0.205$. The dotted line is in the
absence of long-range Coulomb forces ($V_{C}=0$); the solid line
is in the presence of long-range Coulomb forces with $V_{C}=0.55$eV
(adapted from Ref. \cite{BTGD})}
\label{geff}
\end{figure}

In the SB large-N approach, the fluctuations of the 
bosonic fields mediate the residual interaction between the
quasiparticles. If one only considers the fluctuations
of the $b$ and $\lambda$ fields, then only the effects of
the electronic Hubbard repulsion are described and one can
accordingly discuss the effect of the pure electronic
screening preocesses on the e-ph vertex. For the specific
model that we here briefly discuss, Fig. \ref{diagramG}
reports the Feynman-diagram representation of the electronic
processes (schematized by the dashed line of the bosonic
propagators) dressing the bare (empty dot) e-ph vertex.
The ratio between the resulting screened e-ph vertex and the
bare e-ph coupling $g$ is reported in Fig. \ref{geff}
both in the absence (dotted curve) and in the presence (solid curve)
of a long-range Coulomb repulsion. In this latter case the strength is chosen to
produce a repulsion of about $0.1$eV between electrons in nearest-neigbor
cells.
In both cases there is a strong reduction of the static e-ph
coupling. In the short-range case the residual coupling is strongest at
small momenta and it decreases as the momentum.

On the other hand, as it is natural, the long-range potential screens out the long-range charge
fluctuations thereby driving to zero the e-ph coupling at low momenta.

These findings can be obtained and confirmed with several different approaches. In particular, 
they reproduce exactly the results of large-N calculations with Hubbard projectors instead of SB's \cite{KZ1,KZ2}, and are in good
agreement with calculations based on the flow-equation method\cite{aprea} and  recent Gutzwiller+RPA calculations at
finite (but large) U\cite{DLGS}. They also agree qualitatively QMC analysis \cite{Hua03} (although some differences are
present, which can be attributed to the fact that QMC calculations are
performed at finite temperature and at finite Matsubara frequencies)
(cf. Figs. 11 and 12 of Ref. \cite{DLGS}).

All these results show that e-ph scattering at large momenta is
typically weaker than scattering at low momenta in the presence of short range
forces only (which is the case of metallic phases far from phase separation
charge instabilities, where, on the contrary, long-range interctions start to play a crucial
role). This can be of obvious relevance when the relative importance of
e-ph couplings between the quasiparticles and specific phononic modes is
considered \cite{devereaux}. Indeed it might well happen that in the presence
of strong correlations modes that would be strongly coupled, but would exchange
preferably large momenta, are more severly screened than other modes exchanging
lower momenta. All this surely deserves a specific analysis. 

\subsubsection{Dynamical properties}
\label{subsec:dyn}
The previous subsection was focusd on the screening of the static e-ph coupling by 
electronic processes. However the Fermi-liquid analysis carried out in Sec. \ref{sec:sb}
pointed out the great difference between screening processes in the static and in the
dinamical regimes. While the Fermi liquid analysis was only able to provide
definite statement in the small-($q,\omega$) regime, within the SB large-N approach
we can investigate the role of dynamics in the screening processes beyond this limit.

\begin{figure}[t]
\includegraphics[width=7cm,clip=true]{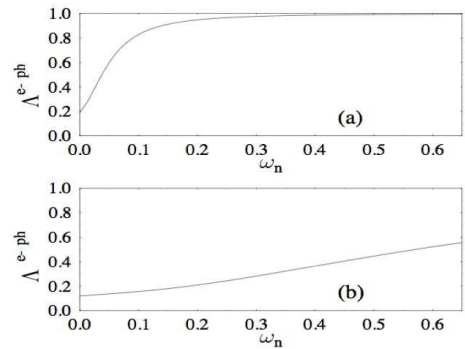}
\caption{Effective e-ph vertex in units of the bare e-ph coupling as a function
of the Matsubara frequency in the HH model at leading order in $1/N$. 
The parameters are the same as in Fig. \ref{geff}. Panel (a) is for a
small transferred momentum ${\bf q}=(0.2,0)$ (in units of inverse lattice spacing
$1/a$); (b) is for a sizable transferred momentum ${\bf q}=(2.0,0)$ 
(after Ref. \cite{BTGD})}
\label{geffs}
\end{figure}

Fig. \ref{geffs} displays the behavior
of the effective e-ph vertex (again normalized to the bare $g$) as a function
of Matsubara frequencies for two distinct momenta in the (1,0) direction for our
HH model in the infinite-$U$ limit.
In panel (a) a small momentum $\qvec =(0.2,0)$ (again unit lattice spacing is used here)
is reported, while panel (b) shows the behavior at a larger momentum $\qvec=(2.0,0)$.
Clearly in the former case the e-ph coupling rapidly returns to its bare value, at
Matsubara frequencies larger than $\sim v_F|\qvec|$ . The $v_F|\qvec|$ scale is instead
much larger in panel (b), where the effective e-ph vertex stays small over a 
much broader frequency range. This fully parallels the low-frequency and low-momentum
limits discussed in Sect. III.A.

Again this result is not specific of this single-band model (or of its treatment)
and it has been confirmed by the analysis of a three-band Hubbard model for the
cuprate CuO$_2$ planes with infinite repulsion on copper orbitals.
In this case, one can even notice in the small-momentum
case an enhancement the effective e-ph coupling above
its bare value. This is an overscreening effect due to interband processes. 
In any case, again one finds that low-momentum processes generically lead to larger
e-ph couplings.

\section{Strong-coupling regime and polaron formation in a correlated metal}
\label{sec:dmft}
\subsection{Proximity to a paramagnetic Mott insulator}
In this section we extend our analysis behind the mean-field level and discuss the fate of the e-ph interaction in a correlated metal under the sole assumption that the physics is governed by the Hubbard repulsion, without assuming a weak e-ph coupling, and/or any approximation as far as the adiabatic ratio is concerned. This means that we need a theoretical approach able to treat severeal energy scales symultaneously, without assuming that any of them is negligible or perturbative.
A natural candidate for this purpose is the Dynamical Mean-Field Theory (DMFT)\cite{revdmft}, which treats all local interaction terms (such as both the Hubbard and the Holstein couplings) and the hopping term on the same footing and it is equally suitable to treat any parameter regime. The central approximation beyond DMFT is the locality of the self-energy (both the electronic and the phonic contributions), a condition which becomes exact when the coordination number becomes large. The original lattice enters in the calculation only through the density of states, which we always choose to be a semicircular one of width $W$.

DMFT allowed to obtain a complete characterization of the Mott-Hubbard transition in the pure Hubbard model, and the emerging physical picture is able to explain several properties of correlated oxides. While we refer to original papers\cite{revdmft} for details, we recall here some aspects which are relevant to our discussion. 

We first consider the situation in which the number of electrons equals the number of lattice sites (the so-called ``half-filling" situation) and, just to focus on pure correlation effects, we consider a paramagnetic phase. In this regime, for large repulsion, the ground state of the HH model can become a ``Mott" insulator, in which the electrons are localized because the electron motion is energetically unfavorable. Starting from the uncorrelated systems and increasing the correlation strength, the spectral weight is transfered from  low to high frequency. In this process the spectral function evolves from a single band to a three-feature structure in which a renormalized band survives around the chemical potential, while precursors of the Hubbard bands develop around $\omega \sim \pm U/2$. As the correlation is further increased up to a value $U \equiv U_{c2}$ the quasiparticle band disappears, and the system becomes a Mott insulator with a preformed Hubbard gap. The key parameter that controls this Mott-Hubbard transition from a metal to an insulator is the quasiparticle weight $z$, which is directly computed from the self-energy (which is a natural outcome of a DMFT calculation). $z$ measures the width and the total spectral weight of the low-energy quasiparticle peak, and its vanishing pinpoints the Mott transition.
As soon as the system is doped away from half-filling, the quasiparticle peak moves towards one of the bands, but it is still well defined for a sizeable doping region (dependent on the value of $U$ and on details of the bandstructure)\cite{revdmft}.
Due to the momentum independence of the self-energy, also in DMFT we have that $z = m/m^*$, so that the Mott transition is associated with a divergent effective mass of the renormalized carriers.

In the following we discuss the effect of a non-perturbative electron-phonon coupling in the strongly correlated metallic solution, i.e., a system in which a quasiparticle peak at the Fermi level is separated from the Hubbard bands realizing a separation of energy scale.
Starting from this situation, in which the quasiparticle bandwidth is $z$ times the bare width $W$, the effect of the e-ph coupling is far from trivial. Indeed there are two main effects associated to e-ph interaction:
\begin{itemize}
\item{The e-ph interaction can introduce a further quasiparticle renormalization, associated to the increase of the quasiparticle effective mass, which may eventually lead to polaronic effects for very strong coupling. This effect leads to a decrease of $z$. In weak coupling we have $1/z = 1 +\frac{\pi}{2} \lambda$ at half-filling and a semicircular density of states of half width $W$.\cite{notalambda}}
\item{The e-ph interaction mediates an attractive density-density interaction (the density-density form is specific for a Holstein coupling), which directly contrasts the Hubbard repulsion. If we integrate out the phonon degrees of freedom, the fermions interact through a dynamical (retarded) interaction. 
\begin{equation}
\label{ueff}
U_{eff}(\omega) = U - \frac{2g^2\omega_0}{\omega_0^2 - \omega^2}.
\end{equation}
In the antiadiabatic regime the frequency dependence of the second term can be neglected and overall interaction is a static term  $U_{stat} = U- \lambda W$, in which the e-ph interaction reduces the strength of the Hubbard term. When the phonon frequency becomes finite the interaction is retarded, but we still expect a similar effect.  Such a decrease of the effective repulsion is expected to make the system less correlated and {\it increase} $z$ (assuming that the repulsion still overcomes the attraction)}
\end{itemize}

The balance between this two effects is not generic and it depends on the adiabatic ratio and on the precise value of the interactions. Yet, important conclusions can be drawn in the correlated regime, in which, also in the presence of e-ph interaction, the separation of energy scales determined by correlations survives.
In this regard, it is important to recall that, within DMFT, the quasiparticle weight is associated to a Kondo resonance of an Anderson-Holstein impurity model. Assuming that the Hubbard $U$ is the largest scale of the problem, the Kondo coupling can be calculated in terms of virtual processes acting in the subspace in which the impurity is singly occupied obtaining an effective Hamiltonian for spin fluctuations \cite{grempel}. The result 
is given by
\begin{equation} \label{grempel1}
J_K(\lambda) = J_K(0) \sum_{m=0}^\infty \frac{ \left\vert \langle 0 \vert \mbox{e}^{g/\omega_0 (a - a^\dagger)} \vert m \rangle  \right\vert^2}{1 - 2 g^2 / \omega_0 U + 2 m \omega_0 / U}
\end{equation}
where the Kondo coupling in the absence of phonon is given by $J_K(0) = 16V^2/U$,   $\vert m \rangle$ is the state with $m$ phonons,  and $V$ is the hybridization between the impurity and the bath. After some algebra, and introducing 
\begin{equation}
U_{eff}=U-\eta \lambda W,
\end{equation}
 we can write, for small $\lambda W/U$
\begin{equation}\label{grempel6}
J_K(\lambda) \simeq J_K(0)\left( 1+\eta\frac{\lambda W}{U}\right) \simeq  \frac{16V^2}{U_{eff}} = \frac{16V^2}{U - \eta \lambda W} 
\end{equation}
with
\begin{equation}\label{eta}
\eta = \frac{2\omega_0 / U}{1+2\omega_0 /U}
\end{equation}
This result would imply that the complicated interplay between the static Coulomb repulsion and the retarded e-ph coupling may be effectively described by an effective purely electronic Hubbard model with a reduced repulsion. We notice in passing that this calculation for the Anderson impurity model is analogous to the evaluation of the effect of phonons on the superexchange coupling of Ref. \cite{superexchange}.
Interestingly the phonon dynamics only enters thtough the ratio $\omega_0/U$, which can be considered as typically small because $U$ is by choice the largest scale of the problem, and $\omega_0$ is smaller than the hopping scale. In the relevant regime of small $\omega_0/U$, we have $U_{eff} = U -4g^2/U^2$.

\begin{figure}[t]
\includegraphics[width=7cm,clip=true]{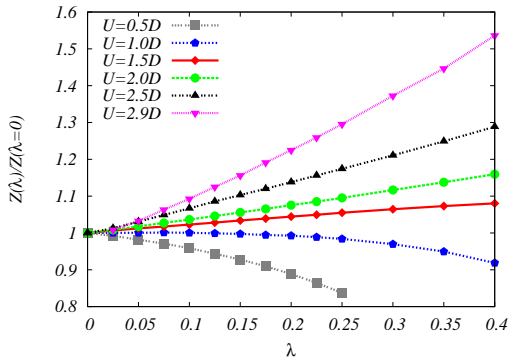}
\caption{Effect of the electron-phonon interaction on the quasiparticle weight $z$ in the presence of electron-electron interaction of different strengths. The quantity plotted as a function of $\lambda$ is the ratio between the full $z$ and the value of the Hubbard model without e-ph coupling ($z(0)$) (after Sangiovanni {\sl et al.} \cite{San05}).
}
\label{fig:Z_lambda}
\end{figure}

These results can be tested through a DMFT solution of the Hubbard-Holstein model at half-filling, where the separation of energy scales is more solid. In Fig. \ref{fig:Z_lambda} we show the ratio $z(\lambda)/z(0)$, in order to emphasize the phonon contribution to the quasiparticle weight. It is apparent that the value of $U$ determines different regimes. For small $U$ the e-ph interaction reduces $z$ (i.e., increases the effective mass) as expected in weakly interacting systems. Increasing $U$ we approach an opposite behavior in which the e-ph interaction makes the quasiparticles {\it lighter} [even if they are obviously heavier than free particles because of the stronger renormalization determined by correlation, which is hidden in $z(0)$].

This behavior confirms that, when the e-e correlation dominates, the leading effect of the e-ph interaction is a reduction of the effective $U$, resulting in an increased quasiparticle mobility. We can now test the above prediction of an effective static repulsion including the effects of e-ph interaction as far as the low-energy physics is concerned.

\begin{figure}[t]
\includegraphics[width=7cm,clip=true]{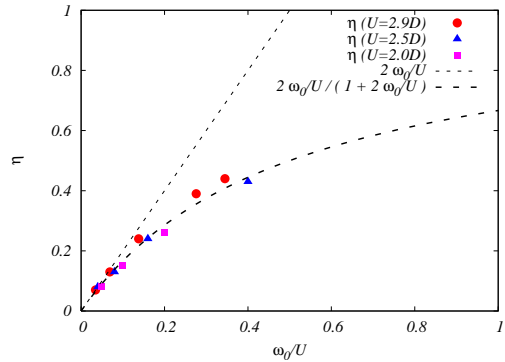}
\caption{Effective static electron-electron interaction for the low-energy properties of the Hubbard-Holstein model. The picture shows the coefficient $\eta$ in $U_{eff} = U -\eta\lambda W$ as a function of $\omega_0/U$ (after Sangiovanni {\sl et al.} \cite{San05}).
}
\label{fig:eta}
\end{figure}

Assuming a form $U_{eff} = U - \eta\lambda W$ for such an effective interaction, we obtained $\eta$ for several values of $U$ and $\omega_0$ simply determining the value of $U$ which gives, for a pure Hubbard model, the same $z$ we obtain for the Hubbard-Holstein model. The results are summarized in Fig. \ref{fig:eta}, and confirm that $\eta$ is essentially a function of $\omega_0/U$ which starts off linear at small values of the argument before bending for larger values. The numerical value beautifully fits the value given above on the basis of the analogy with Kondo effect. 

\begin{figure}[t]
\includegraphics[width=7cm,clip=true]{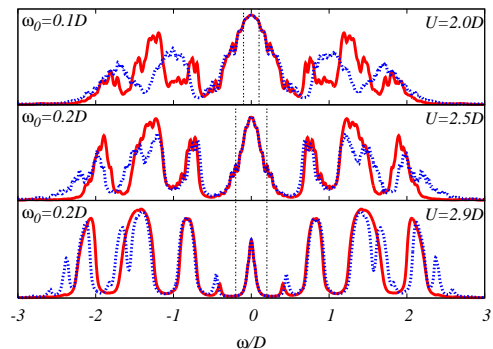}
\caption{Comparison between the DMFT spectra for the Hubbard-Holstein model and the effective purely electronic Hubbard model with $U = U_{eff}$. The low-energy part of the spectra is perfectly reproduced by the effective model, while the high-energy features present phononic signatures that can not be plugged in an effective static interaction (after Sangiovanni {\sl et al.} \cite{San05}).
}
\label{fig:spectra}
\end{figure}

On the basis of our knowledge about the DMFT results for the pure Hubbard model, the quasiparticle weight completely characterizes the low-energy quasiparticle peak. Therefore our analysis should imply that the low-energy part of the spectral function of our Hubbard-Holstein model can be described by means of the effective Hubbard model that we introduced. This is strikingly confirmed by a direct comparison, as shown in Fig. \ref{fig:spectra}. Here we show some examples of spectra for the Hubbard-Holstein model  compared with the associated effective Hubbard model. 
Besides the spectacular confirmation of the validity of the effective model for the low-energy part of the spectrum, another important information emerges from the picture: The high-energy part of the spectrum is instead affected by phonons in a more ``dynamical" way, meaning that the high-energy Hubbard bands acquire a modulation in frequency which can be related to phonon satellites, which are completely absent in the low-energy part, where nothing happens at the characteristic phonon frequency. 

The picture that emerges from the previous arguments can be summarized as follows: Quasiparticle motion arises from virtual processes in which 
doubly occupied sites are created. Obviously, these processes are not so frequent, since the energy scale involved is large, but they are 
extremely rapid (the associated time scale is $\propto 1/U$), and consequently are poorly affected by phonon
excitations with a characteristic time scale $1/\omega_0 \gg 1/U$.  When the phonon frequency is small with respect to $U$, 
the phonon degrees are frozen during the virtual excitation processes.
Therefore, despite the overall electron motion is quite slow due to the small number of virtual processes 
(which is reflected by the large effective mass), the e-ph interaction has no major effect except for a slight reduction of the
total static repulsion.

According to what we have just described, we can conclude that strong correlations reduce the effect of the e-ph interaction on the low-energy properties, associated to quasiparticle propagation, while the high-energy properties present more standard phonon signatures, such as the satellites at the phonon-frequency scale.

It has to be underlined that DMFT is not able to introduce momentum-dependent corrections to the electronic properties. The above analysis therefore  shows indeed that the ``standard" electron-phonon interaction is heavily screened (and it actually loses its dynamical nature) when the low-energy quasiparticle properties are considered. Only ``non-standard" effects, such as the prevalence of forward scattering that we discussed in Sec. \ref{sec:fl} can survive at low energy.

The above scenario has been carried out at half-filling, where the presence of e-e correlations has its most striking effects, both in terms of the phase diagram, as it can give rise to a Mott transition, and in terms of the separation of energy scales, which is clearly sharper than for doped systems.
Therefore, as soon as we dope the Hubbard-Holstein model, even for $U > \lambda$ the dominance of repulsive correlations is weaker and the interplay with e-ph coupling will be more subtle\cite{San06}. As a result, it is not possible to describe the effect of phonons on the highly correlated metal in terms of an effective static potential unless the system is very close to the antiadiabatic limit. Therefore, for small and intermediate phonon frequencies, we  do not find situations in which increasing the electron-phonon coupling reduces the effective mass.

\begin{figure}[t]
\includegraphics[width=7cm,clip=true]{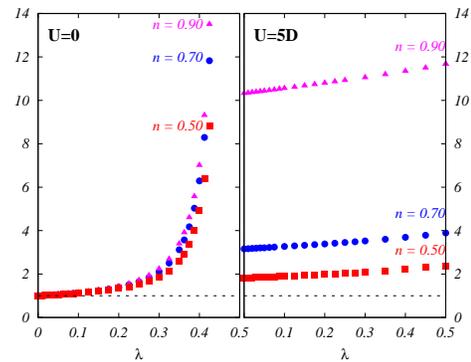}
\caption{Effective mass $m^*$ as a function of $\lambda$ for three different densities different from half-filling (n=0,5, 0.7, 0.9) in the uncorrelated (left) and strongly correlated ($U=2.5W$, right) system (after Sangiovanni {\it et al.} \cite{San06})}
\label{fig:doping1}
\end{figure}

Nonetheless, if we consider reasonably large  values of $U$, the dominance of correlations will determine a reduce effectiveness of the e-ph coupling, and, e.g., polaron formation will be pushed to significantly larger values of $\lambda$ then for uncorrelated systems, as shown by the DMFT results of Fig.\ref{fig:doping1}. Here we plot $m^*/m$ as a function of $\lambda$ and we compare the uncorrelated system with the system with $U/W$=2.5 (Here, since the density is different from half-filling, the system is always metallic even if $U$ is larger than the critical value for the Mott transition). While in the uncorrelated case $m^*$ grows exponentially when $\lambda$ approaches a critical value of order 1 for all densities, signaling polaron formation, the correlated system displays a significantly weaker growth of $m^*$ up to $\lambda \simeq 0.5 \div 0.75$. 

\begin{figure}[t]
\includegraphics[width=7cm,clip=true]{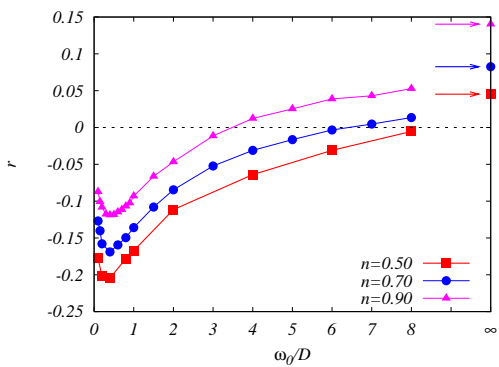}
\caption{Renormalization of the quasiparticle electron-phonon coupling as a function of $\omega_0/W$ for three different densities and $U/W=2.5$ (after Sangiovanni {\it et al.} \cite{San06})}
\label{fig:doping2}
\end{figure}

The role of the adiabatic ratio is illustrated by Fig. \ref{fig:doping2}, where we report the renormalization of the linear coefficient of the mass enhancement defined by the relation $m^*(U)/m^*(U,\lambda) = 1 -r\lambda$. Here $m^*(U,\lambda)$ is the effective mass in the presence of both electron-electron and electron-phonon interactions and $m^*(U)$ is the same quantity in the absence of coupling to the phonons. Here a negative $r$ implies a standard increase of the effective mass due to phonons. The results (again for $U/W$=2.5) show that in all cases the coefficient is smaller than one, confirming that correlations reduce the effective e-ph coupling, and a strong (and nonmonotonic) dependence on the antiadiabatic ratio. Only for very large values of $\omega_0/W$ $r$ becomes positive reflecting that the ``screening" physics we described above is effective. The dependence on the density naturally reflects that fillings closer to $n=1$ display weaker phononic effects in the adiabatic regime of small frequencies and a more rapid evolution to an antiadiabatic regime in which $r > 0$.

\begin{figure}[t]
\includegraphics[width=7cm,clip=true]{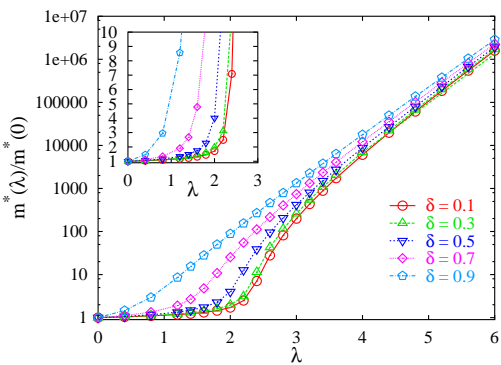}
\caption{Renormalization of the effective mass due to electron-phonon
coupling for infinite $U$ and $\omega_0/W = 0.1$ as a function of
$\lambda$ for different doping levels. The inset shows the same data in
a smaller range. Here $\lambda$ is twice as in the rest of the paper. (after Barone {\it et al.} \cite{barone1})}
\label{fig:gutz1}
\end{figure}

The scenario which emerges from DMFT calculations at finite $U/W$ can be confirmed by a semi-analytical approach based on an extension of the Gutzwiller approach which treats phonon degrees of freedom on the same footing of the electrons. For more details see Refs. \cite{barone1,barone2}.
The results of Fig. \ref{fig:gutz1} in the limit of infinite repulsion confirm the scenario arising from Fig. \ref{fig:doping1}: in the correlated system the electron-phonon interaction is considerably reduced with respect to the free system, but the qualitative effect of $\lambda$ is an increase of the effective mass.

\subsection{Antiferromagnetic correlations}
The above analysis suggests that, as expected, strong e-e correlation essentially opposes to e-ph coupling, eventhough ``anomalous" signatures of e-ph coupling still survive even in regions of parameters for which correlations prevail. Yet, these results are limited to the metallic paramagnetic state, in which no broken symmetry is allowed. At half-filling and for some finite doping region, strong correlations lead to an antiferromagnetically ordered state, and it is expected that finite-range antiferromagnetic correlations survive in a wider doping region. The relation between antiferromagnetism and e-ph interaction is hinted by the experimental framework. Indications of phonon signatures in high-T$_c$ superconductors are indeed particularly strong in the extremely underdoped region, where some kind of antiferromagnetic correlation is certainly present. For example, clear polaronic features are observed in the optical spectroscopy\cite{optics_polarons} and ARPES\cite{KMShen} of underdoped materials. 

From a theoretical point of view, several investigations indeed suggest that the e-ph interaction is particularly effective for a hole in an antiferromagnetic background\cite{hole,napoli}, and for slightly doped t-J models. These results can be reconciled with the above findings for the nonmagnetic phase by simple arguments.

Within the paramagnetic phase, the effect of increasing correlations is a strong reduction of the quasiparticle weight $z$ associated with a divergent self-energy, which in turn strongly renormalizes the e-ph vertex, leading to the strong reduction of low-energy phononic signatures that we described above. Once antiferromagnetic correlations are allowed, the system can turn insulating even with a finite $z$ and a non-divergent self-energy, hence the e-ph vertex is not severely screened. 
From a more physical point of view, the antiferromagnetic insulator allows for more charge fluctuations with respect to the pure Mott state at the same value of $U$. Therefore a Holstein coupling, which exploits precisely charge fluctuations to gain energy, is expected to be favoured by antiferromagnetic correlations.

\begin{figure}[t]
\includegraphics[width=7cm,clip=true]{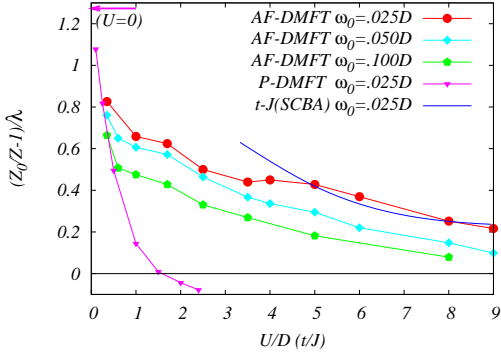}
\caption{Effect of the Coulomb repulsion on the linear contribution in $\lambda$ to the quasiparticle renormalization. DMFT results for the antiferromagnetic phase (AF-DMFT) are compared with DMFT in the paramagnetic state (P-DMFT) and to a self-consistent Born Approximation (SCBA) for the t-J model for parameters corresponding to the red dots AF-DMFT. Antiferromagnetism allows for a sizeable e-ph coupling also in the present of very strong Coulomb repulsion, in constrast with paramagnetic results (after Sangiovanni {\it et al.} \cite{Sangiovanni_AFM})}
\label{fig:zz0_afm}
\end{figure}

Direct DMFT calculations in the antiferromagnetic phase at half-filling confirm these expectations. In Fig. \ref{fig:zz0_afm} we show the quantity
$[z/z(0)-1]/\lambda$ for small $\lambda$ ($z(0)$ being $z$ in the absence of e-ph interaction). This quantity measures the renormalizaton of the linear e-ph coupling induced by e-e correlations. The comparison between the paramagnetic solution and the antiferromagnetic state confirms the above expectations. While this coefficient rapidly drops as a function of $U$ in the paramagnetic state, the inclusion of antiferromagnetism leads to a much more robust e-ph coupling. We underline that, however, the e-ph interaction is still substantially reduced with respect to the non-interacting systems. 

\begin{figure}[t]
\includegraphics[width=7cm,clip=true]{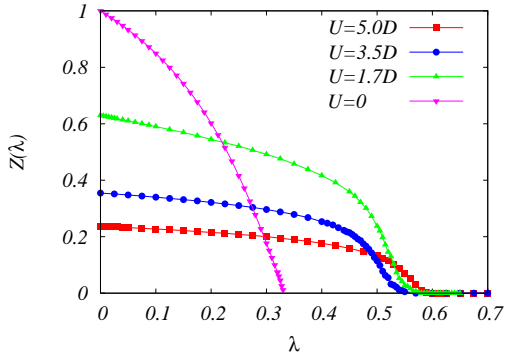}
\caption{Quasiparticle renormalization as a function of $\lambda$ for different values of $U$ and $\omega_0 =0.0125W$. The critical value for polaron formation is only moderately increased by correlations with respect to the non-interacting case (after Sangiovanni {\it et al.} \cite{Sangiovanni_AFM})}
\label{fig:z_afm}
\end{figure}

The comparison with the uncorrelated system is shown in Fig. \ref{fig:z_afm}, where the evolution of $z$ as a function of $\lambda$ is followed beyond the perturbative regime. In all cases $z$ decreases monotonically until it vanishes signaling polaron formation. Yet, the decrease is more rapid for the uncorrelated system, and the critical coupling is significantly enhanced in the correlated antiferromagnetic state.

\section{Phonon mediated charge instabilities}

In the previous sections we focused on the metallic phases, without discussing the possible instabilities,
either directly driven by the interaction terms, or favoured by the weakness of the correlated metallic state.
One can indeed expect, on very general grounds, that the reduced kinetic energy characteristic of the
strongly correlated metal can be easily overcome by different localizing effects thereby destabilizing the metal in favor of ordered phases. 

As we did in Sec. \ref{sec:fl} in the discussion of the e-e screening of the e-ph vertex, we can start our analysis in the FL framework before discussing in some more details
model-specific results.

In particular, we focus on charge instabilities, which have been invoked to play an important role in high-$T_c$ 
superconductors. 
The charge compressibility $\kappa = \partial n/\partial\mu$, which in a FL theory reads
\begin{equation}
\kappa = \frac{N^*}{1+F_0^s}
\end{equation}
is the key quantity that controls the stability of the charge degrees of freedom. A positive $\kappa$ is the stability 
condition. Therefore if $F_0^s < -1$ this condition is violated and the system undergoes a charge instability towards
a phase-separated state.

It is important to stress that the {s of Sec. \ref{sec:fl} about the reduction of the static e-ph vertex do not 
necessarily imply that e-ph effects do not  contribute to the charge compressibility in the presence of strong correlations.
Indeed $F_0^s\equiv 2 N^* \Gamma_\omega$, where $\Gamma_{\omega}$ is the full {\it dynamic} effective scattering
amplitude between the quasiparticles\cite{Nozieres} including both e-e and e-ph interactions. As opposed to its static counterpart,
the dynamic amplitude which enters $F_0^s$ is not depressed by the e-e vertex corrections.
For example, at lowest order in $g^2$ and by performing an infinite-order RPA resummation of the e-ph screening processes,
$F_0^s$ reads
\beq
F_0^s=2N^*\left(  \Gamma_\omega^e -g^2\right),
\eeq
where $\Gamma_{\omega}^e$ is the dynamic vertex function determined by e-e correlation processes only.
This equation indicates that a sufficiently large bare $g^2$ can overcome the effective residual repulsion between the 
quasiparticles $\Gamma_\omega$ leading to a phase separation instability marked by the 
Pomeranchuk condition $F_0^s<-1$. It is worth pointing out that $m^*/m\gg 1$ requires a large
{\it bare} repulsion between the physical electrons (a large Hubbard $U$ in our model) but this by no means requires a
large residual  interaction between the heavy quasiparticles. 
Therefore even a small e-ph interaction can give rise to a phase separation instability.

%The lowest-order analysis 
%showing the depression of the e-ph coupling in the low-energy processes maintains its full
%validity with respect to the inclusion of higher-order terms in $g$ provided $F_0^s$
%in Eq. (\ref{geffoneph}) is positive and still of order $m^*/m$. 
Moreover, near the instability 
condition $F_0^s=-1$, the phonon contribution to the vertex becomes substantial and the e-ph
interaction becomes relevant even in the static limit, at least at small $q$'s. At large $q$
the analysis of specific models shows that the e-e interaction mediated by phonons is instead
suppressed also near the instability region.

This is what indeed happens in the HH model treated within the SB-large-N method described
In Sec. \ref{sec:sb}. Within this approach it was first demonstrated that a metal with moderate e-ph coupling and strong e-e
correlations could undergo a charge instability \cite{CDG,BTGD}.
Specifically in the absence of long-range Coulombic forces the Pomeranchuk stability condition is violated. The doped 
HH model does not form a uniform phase and the system undergoes a phase separation between hole-rich regions and insulating half-filled
regions.

 %Within the usual Fermi liquid description, this phase separation instability
%was signaled by  a divergent compressibility
%\begin{equation}
%\kappa\equiv \frac{\partial n}{\partial \mu}=
% \nu^* (1-\nu^* \Gamma_q)=\frac{N\nu^*}{1+F_0^s }{\to_{F_0^s\to -1} \infty}
%. \label{compressibility}
%\end{equation}
%when the Pomeranchuk criterion for stability $F_0^s>-1$ is violated.
In the presence of long-range Coulomb interactions the electrostatic cost of the charge-rich regions
would become infinite, and the thermodynamic phase separation can not take place.
The tendency towards charge segregation would however survive and it finds its realization
with a charge-density wave instability which emerges as a compromise between the charge
segregation tendency and the homogeneizing effect of long-range interactions.
This  mechanism for charge ordering is the so-called ``frustrated
phase separation'' \cite{frustratedPS,RCGBK,low94,LCD}.
For the specific HH model described here it was found that for
realistic values of the e-ph coupling and of the long-range repulsion, frustrated phase
separation gives rise to 
a second-order quantum phase transition (quantum critical point, QCP) around optimal doping 
(doping $x=0.19$) \cite{CDG} with the ordering and the periodicity influenced but
not directly related to the structure of the Fermi surface.
This instability arises instead from the energetic balance between the
tendency to phase separation and the frustrating electrostatic cost of the
long-range Coulomb interaction. Near this instability the phonon spectrum becomes highly anomalous.
First of all the phonon acquires a strong coupling to the electronic degrees of
freedom near the instability wavevector $\qvec_c$ [which usually tend to occur
at $\qvec_c\approx (\pm \pi/2,0), (0,\pm \pi/2)$ for the relevant dopings
and Fermi-surface shapes]. Near the instability wavevector the phonon
linewidth becomes therefore very broad and it even acquires a background
of the order of the particle-hole continuum. At the same time the phonon
dispersion softens and at the instability the frequency of this mixed 
phonon-electron mode vanishes. Remarkably, since the critical wavevector is
not so large (typically of the order of $\pi/2$) the region where the
phonon dispersion becomes strongly anomalous is rather isotropic and
substantial anomalies are present also in the (1,1) direction upon
approaching the critical doping of the QCP. Fig. \ref{phonfreq} reports 
the anomalous phonon dispersion found in the HH model in Ref. \cite{BTGD}.
\begin{figure}[t]
\includegraphics[scale=0.3,angle=-90,clip=true]{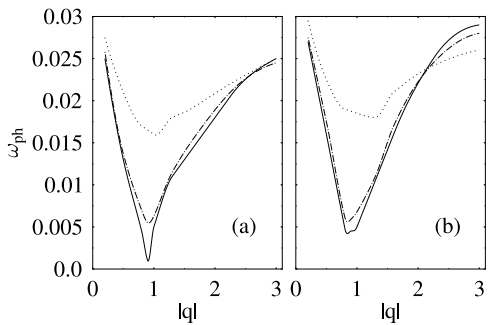}
\caption{Phonon dispersion curves (a) in the 
instability (0.28,0.86) direction and (b) in the (1,1) direction, 
for $t=0.5$eV, $t'=-(1/6)t$, $g=
0.240/\sqrt{2}$eV, $\omega_0=0.04$eV, and 
$V_{C}=0.55$eV. The solid curves
correspond to the critical doping $\delta_c=0.195$, the 
dot-dashed and the dotted curves correspond to 
$\delta=0.21$ and $\delta=0.3$ respectively.
}
\label{phonfreq}
\end{figure}

Of course it is quite tempting to relate these anomalies, to the anomalies 
observed by inelastic neutron scattering \cite{mcqueeney,pintschovius,reznik}.
An alternative possibility can also be proposed for the anomalied detected
in underdoped cuprates:
charge ordering can give rise to rather anisotropic 
nearly one-dimensional dynamical charge textures. In this case Kohn anomalies
can be expected along the stripes at wavevectors of the order $2k_F$ in the
stripe direction \cite{diclorenz}

\section{Jahn-Teller coupling in the Fullerenes}

All the above analysis has been carried out for the HH model. It has to be emphasized that some
of the effects we discussed may be less general than what the simple form of the Hamiltonian may suggest.
As we discussed in details, the HH model is indeed characterized by two interaction terms which are both
related to the charge degrees of freedom, and they indeed directly compete, as clearly shown by Eq. (\ref{ueff}).
This direct competition makes the two effects more exclusive than in general situations in which the e-ph coupling
does not directly compete with the Coulomb repulsion. We can have two different ways to avoid the direct competition:
a different functional form for the e-ph coupling in a single-band model, or a Jahn-Teller couping in a multiorbital model.

The first situation can obviously have relevance for the cuprates, in which different phonon modes with specific smmetries
may play a role, or for system dominated by the so-called Su-Schrieffer-Heeger coupling in which the phonons modulate
the nearest-neighbor hopping. The second situation occurs instead in the fullerenes, where the relevant conduction band
is a three-fold degenerate manifold of $t_{1u}$ ($p$-like) orbitals, which couple with Jahn-Teller active local distortions 
of the fullerene molecule. It is precisely this kind of coupling which is expected to be responsible for supercondutivity in these
compounds\cite{Gunnarsson_C60}.

A three-band model which includes a strong Coulomb repulsion, a Hund's rule splitting and a moderate Jahn-Teller e-ph coupling
has been studied in several papers\cite{Capone_C60_1,capone_science,Han03,capone_rmp}, reaching an {\it a priori} suprising concousion: The Jahn-Teller coupling is not harmed by large Coulomb repulsion, and the phonon-driven superconductivity can actually be strongly enhanced in the proximity of the Mott transition, i.e., in the region in which
the correlations are most effective.

Here we do not discuss the physics of this model in details, since we are mainly interested in contrasting its behavior with the HH model. 
The key point is that the Jahn-Teller interaction does not touch the total charge on each molecule, as it couples with a combination of 
local spin and orbital degrees of freedom. As a consequence, even when correlations are sufficiently strong to suppress the electronic
motion, the localized electrons can still interact within a single fullerene molecule via the e-ph interaction. For example, if we consider the experimentally relevant
situation of three electrons per fullerene molecule, as we approach the Mott state three electrons will remain stuck on each fullerene. 
Yet, the can still be in a high-spin state or low-spin state, and the energetic gain associated to the multiplet splitting will the the same as
for a non-interacting molecule. Therefore, the e-ph driven interaction will be not renormalized by correlations, as opposed to the Holstein model.
From a Fermi-liquid point of view, the lack of renormalization is determined by very large vertex corrections (divegent like $1/z$ as the Mott 
transition is approached) that compensate the $z$ factors.

As a matter of fact, the effective interaction between quasiparticles obtained in a nonperturbative DMFT study of the model corresponds to a 
severely screened Hubbard repulsion plus an essentially unscreened phonon-driven attraction that can be parameterized as 
\begin{equation}
A_{eff} = zU - \frac{10}{3}J.
\end{equation}
This simple equation shows that, even if $U$ is chosen to be significantly larger than $J$, when the Mott transition is approached (i.e., $z\to 0$)\cite{Capone_C60_1}, the 
attraction will eventually prevail. Moreover, in this regime the quasiparticles are quite heavy, and their large effective density of states can lead to an enhancement of the effective dimensionless coupling, which is expected to reflect in an enhanced critical temperature. 

This is precisely what is found by explicitly solving the superconducting phase of our three-band model in DMFT. If we follow the evolution of the superconducting order parameter as a function of $U$ for a fixed small $J$ we first have a standard BCS-like region when $U$ is so small that the bare attraction simply overcomes the bare repulsion. Then superconductivity disappears because $U$ is large enough to kill the attraction, but $z$ is still close to 1. Further increasing $U$ we approach the Mott transition, $z$ decreases and it strongly renormalizes down the effective repulsion. Eventually the effective interaction becomes attractive and superconductivity re-emerges, with an order parameter that  follows a bell-shaped curve before vanishing at the Mott transition point. This strongly correlated superconducting pocket displays a maximum critical temperature which exceeds the weak-coupling BCS value. In other words, phonon-driven superconductivity is actually enhanced by strong correlations\cite{capone_science}. 

A full DMFT solution of the model has allowed both to predict the experimental observation only later provided in Refs. \cite{C60_corr}, like the dome-behavior of the critical temperature as a function of doping, and the first-order transition to a spin-1/2 antiferromagnet when pressure is reduced to recover the ambient phase of A15 Cs$_3$C$_{60}$, and to further characterize the properties of strongly correlated superconductors. For example, we predict a pseudogap in the photoemission spectra, and a kinetic-energy driven superconductivity for the most expanded compounds\cite{capone_rmp}.

In the context of this review, our solution is a clear example of the crucial role of the phonon symmetry. In our multiband model it is possible to consider phonons which are by symmetry unharmed by correlations, as opposed to the Holstein model. The result is confirmed by investigations of simplified two-orbital models which share the same properties\cite{Capone_C60_2,SchiroPRB} 
When we go back to the cuprates, and to single-band models, our findings suggest that phonon modes which are coupled to operators which are not proportional to the charge can indeed survive much better in a strongly correlated environment, in analogy with the findings of mean-field methods.

\section{Conclusions}

The focus of the present topical review is on the effects of strong electron-electron correlations on the electron-phonon coupling.
We mostly considered the Hubbard-Holstein model, where both e-e and e-ph interactions locally couple to electron density
fluctuations. In this case the competition between these interactions is quite effective, particularly
in the proximity of the Mott-Hubbard transition. Indeed, since
the e-e Hubbard repulsion makes the density fluctuations stiffer, the Holstein e-ph coupling $g$ is generically
suppressed. Along this paper we made more specific this point by reviewing the specific momentum and frequency
dependencies of this suppression. A first observation is that the suppression is strong whenever the quasiparticle
residuum $z$ is small. On the contrary we observed that in the antiferromagnetic phase, where the suppression
of double occupancy imposed by the large Hubbard repulsion $U$ is due to the spin ordering and does not entails
a small $z$, the e-ph coupling is only weakly suppressed. This easily explains why polaronic features are
present and clearly visible in weakly doped antiferromagnetic cuprates. The weakness of this suppression
is likely to persist even in the metallic paramagnetic regime, if substantial residual antiferromagnetic correlations
persist on a local basis. The opposite case of a strong suppression of the e-ph coupling ,occurring for small $z$ (or more precisely when
$\kappa^e/N^* \ll 1$), needs
further specification. In particular we find that $g$ is more or less suppressed depending on the dynamical
regime: for a small ratio between the transferred frequency $\omega$ and the transferred momentum $v_F q$
the e-ph coupling is strongly reduced, while in the opposite limit $\omega/v_Fq \gg 1$ no suppression is found
and even an enhancement is possible.
This latter finding leaves the possibility open of substantial phononic residual attractions between
the quasiparticles competing with the residual repulsions in driving the system unstable toward long-wavelength
charge instabilities. These two dynamical regimes are also visible in the DMFT approach where the scale $v_Fq$ is reflected in the width of
the quasiparticle resonance. For frequencies smaller than this latter scale phononic features are absent, while they are
clearly present at high-energy in the Hubbard side-bands.

The nearly static case $\omega/v_Fq \ll 1$ again displays an intrinsic richness as far as
momentum dependence is concerned: the strong suppression
of $g$ already occurring at small momenta becomes really very strong at large transferred momenta. This suppression
found both on general grounds within a Fermi-liquid analysis and within specific field-theoretic treatments
of the Hubbard-Holstein model \cite{GC,BTGD,KZ1,KZ2,aprea} can account for the
impressive elusiveness of phononic features in transport experiments in cuprates. Indeed the fact that the resistivity in the
metallic phase does not display any clear phonon-related feature is naturally  explained by the strong suppression
of the e-ph coupling in transport processes, where very low-energy and large transferred momenta are involved.
Thus the strongly correlated nature of the cuprates is the key ingredient to solve
the puzzles related to the dichotomic behavior of these materials, which display clear phononic features
in some cases and none in others. 

All the above findings are only partly peculiar of the specific HH model, where the electron density locally 
involves both the e-e and e-ph coupling: The analysis of other models like the SSH essentially produce the
same results \cite{KT,ODLSG}. A qualitative difference only occurs for those phonons which couple to degrees of freedom,
which are not severely stiffened by the proximity to a Mott-Hubbard phase. In this regard we reported
the important case of Jahn-Teller phonons in the Fullerenes. It would be interesting to search for
similar phononic (or even non-phononic) degrees of freedom in the cuprates. In this regard, the buckling modes
in some cuprates are interesting candidates, which are presently being investigated in this perspective
\cite{SLG}.

\section*{Acknowledgments}
The work that we review has been carried out in collaboration with G. Sangiovanni, P. Barone, S, Ciuchi, C. Di Castro, J. Lorenzana, G. Seibold, A. Di Ciolo, M. Fabrizio, E. Koch, O. Gunnarsson, P. Paci, R. Raimondi, and A. Toschi. We acknowledge financial support by MIUR PRIN 2007 Prot. 2007FW3MJX003. M.C.'s activity is funded by the European Research Council under FP7/ERC Starting Independent Research Grant ``SUPERBAD" (Grant Agreement n. 240524).  M. G. acknowledges financial support from the Vigoni Program of the Ateneo Italo-Tedesco.

%Unused bibitems

\begin{thebibliography}{99}
\bibitem{notalambda} Here and in the following we define $\lambda = \frac{2g^2}{\omega_0 W}$ replacing the density of states at the Fermi level with $1/W$.
\end{thebibliography}
\end{document}